\documentclass[amsmath,amssymb,aps,pra,twocolumn]{revtex4-1}

\usepackage{graphicx}
\usepackage{dcolumn}
\usepackage{bm}

\newcommand{\be}{\begin{equation}}
\newcommand{\ee}{\end{equation}}
\newcommand{\ben}{\begin{eqnarray}}
\newcommand{\een}{\end{eqnarray}}

\begin{document}
\title{Generation of High-Order Vortex  States from Two-Mode Squeezed States}

\author{Graciana Puentes$^{1,2}$ and Anindya Banerji$^{3}$   }

  \affiliation{1-Departamento de Fsica, Facultad de Ciencias Exactas y Naturales, Universidad de Buenos Aires, Ciudad Universitaria, 1428 Buenos Aires, Argentina\\ 
   2-CONICET-Universidad de Buenos Aires, Instituto de Fsica de Buenos Aires (IFIBA), Ciudad Universitaria, 1428
Buenos Aires, Argentina. \\
3-Centre for Quantum Technologies, National University of Singapore, Singapore.}

\date{\today} 

\begin{abstract}
We report a scheme for generation of high-order vortex states using two-mode photon-number squeezed states, generated via the non-linear process of Spontaneous Parametric Down Conversion. 
By applying a parametric rotation in quadrature space $(X,Y)$, using a $\phi$ converter, the Gaussian quadrature profile of the photon-number squeezed input state can be mapped into a superposition of Laguerre-Gauss modes with $N$ vortices or singularities,
for an input state containing $N$ photons, thus mapping photon-number fluctuations to interference effects in quadrature space. Our scheme has the potential to improve measurement sensitivity beyond the Standard Quantum Limit ( SQL $\propto \sqrt{N}$), by exploiting the advantages of optical vortices, such as ease of creation and detection, high dimensionality or topological properties, for applications requiring reduced uncertainty, such as quantum cryptography, quantum metrology and sensing. 
\end{abstract}


\maketitle

\maketitle

\section{Introduction}

In quantum optics, a beam of light is in a squeezed state if its electric field amplitude has a reduced uncertainty, in relation to that of a coherent state. Thus, the term squeezing refers to squeezed uncertainty. In general, for a classical coherent state with $N$ particles, the sensitivity of a measurement is limited by shot noise to the Standard Quantum Limit (SQL $\propto \sqrt{N}$).
On the other hand, quantum states, such as photon-number squeezed states, hold the promise of improving measurement precision beyond the SQL. Squeezed states of light find a myriad of applications, such as in precision measurements, radiometry, calibration of quantum efficiencies, or entanglement-based quantum cryptography, to mention only a few [1-10]. \\

An optical vortex is a singularity or zero point intensity of an optical field. More specific, a generic Laguerre-Gauss beam of order $m$ of the form $\psi \propto e^{ i m \phi}e^{ -r^2}$, with $\phi$ its azymuthal phase and $r=\sqrt{x^2+y^2}$ its radial coordinate, has an optical vortex in its center for $m>0$. The phase in the field circulates around such singularity giving rise to vortices. Integrating around a path enclosing a vortex yields an integer number, multiple of $\pi$. 
This integer is known as the topological charge. There is a broad range of applications of optical vortices in diverse areas, such as in astronomy for detection of extra-solar planets, in optical tweezers for manipulation of cells and micro-particles, 
in optical communication to improve the spectral efficiency, in Orbital Angular Momentum (OAM) multiplexing, and in quantum cryptography to increase communication bandwidth [11-20]. \\

In this article, we report a scheme for generation of high-order vortex states using two-mode photon-number squeezed states generated via the non-linear process of Spontaneous Parametric Down Conversion (SPDC). 
By applying a parametric rotation in quadrature space $(X,Y)$ using a $\phi$ converter, the quadrature representation of the photon-number squeezed input state can be mapped into a vortex state in quadrature space containing $N$ vortices or singularities,
for an input state containing $N$ photons, thus mapping photon-number fluctuations to interference effects in quadrature space, giving rise to thee emergence of a state with a well-defined number of vortices. 
Our scheme has the potential of exploiting the advantages of optical vortices, such as ease of creation and detection, high dimensionality or topological properties, for applications requiring precision beyond the SQL $\propto \sqrt{N}$, such as quantum cryptography, quantum metrology and sensing.\\
The article is structured as follows: First, in Section II we review the properties of two-mode photon-number squeezed states such as their quadrature representation and photon-number distribution, second in Section III we introduce the concept of quadrature rotation. Next, in Section IV, we present the spatial quadrature representation of the rotated states in terms of Laguerre-Gauss modes. In Section V, we present numerical simulations confirming the creation of high-order vortex states in quadrature space containing $N$ vortices for a squeezed input state containing $N$ photons. In Section VI, we present analytical and numerical derivations for the photon-number distribution of the resulting vortex states, revealing super-Poissonian photon statistics. Finally, in Section VII, we present our conclusions. 
\section{2-mode photon-number squeezed state}
\begin{figure}[t]
\hspace{-0.5cm}
\includegraphics[width=0.5\textwidth]{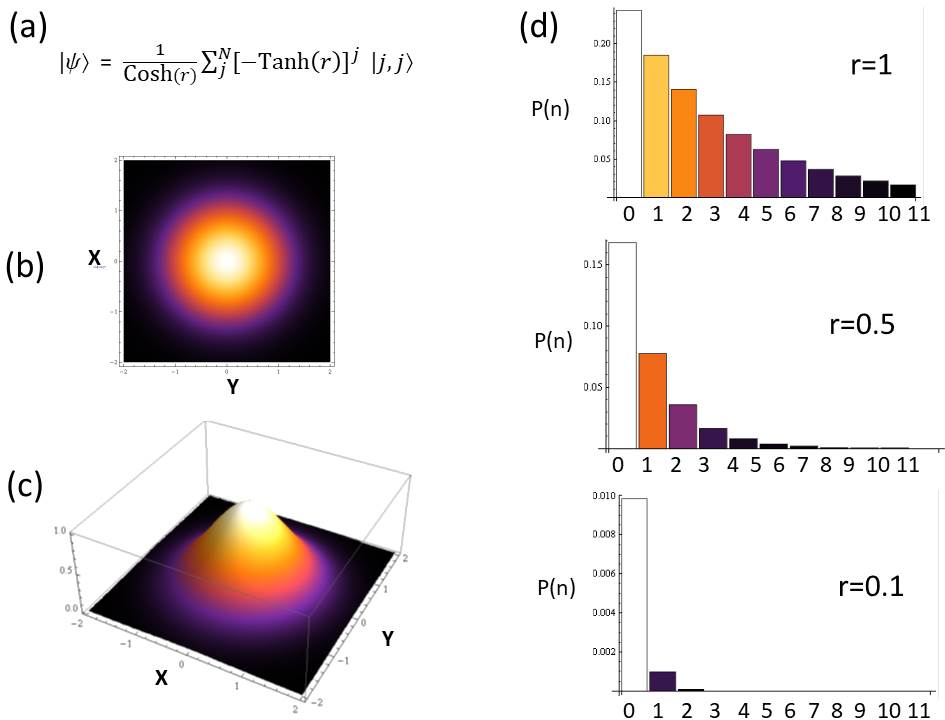}
\caption{(a) Two-mode photon-number squeezed input state $|\psi \rangle$. (b) and (c) Spatial quadrature representation of input state $\psi(x,y)$ displaying a Gaussian spatial profile with no vortices or singularities. (d) photon-number distribution for different values of squeezing parameter $r=1, 0.5, 0.1$, revealing sub-Poissonian quantum statistics (see text for details). }
\end{figure}

Consider a truncated two-mode photon-number squeezed state with $N$ photons in the Fock state representation of the form \cite{Caves}:
\begin{equation}
|\psi \rangle =\frac{D}{\cosh {r}} \sum_{j=0} ^{N/2} (\tanh{r})^{j}|j,j\rangle,
\end{equation}
where $D$ is a nornalization factor and $r$ is the squeezing parameter. The spatial representation of the photon-number states $| n_{x}, n_{y}\rangle$  can be written in terms of the stationary solutions of the Harmonic Oscillator, which are customarily expressed as a product of Hermite-Gauss polynomials of the form $\psi(x,y)_{HO}=\sqrt{\frac{1}{\pi 2^{(n_{x}+n_{y})} n_{x}!n_{y}!} } H_{n_{x}}(x) H_{n_{y}}(y) e^{-(x^2+y^2)/2}$ \cite{Karimi}. Using such expressions, the  photon-number squeezed state has a spatial representation in quadrature space $(X,Y)$ of the form $ \psi(x,y)=\langle x,y |\psi \rangle$:
\begin{equation}
\psi(x,y)=\frac{D}{\cosh {r}} \sum_{j=0} ^{N/2} (\tanh{r})^{j} H_{j}(x) H_{j}(y) e^{-(x^2+y^2)2},
\end{equation}
where $ H_{j}(x) = \langle x| j\rangle$ and $ H_{j}(y) = \langle y| j\rangle$ correspond to Hermite-Gauss polynomials of order $j$ in quadrature space \cite{Karimi}. The quadrature representation of the input state $|\psi\rangle$ reveals a classical Gaussian profile, with no vortices or singularities. 
This is depicted in Fig. 1 (b) and Fig. 1 (c). Therefore, the squeezing in uncertainty for the input state is in the photon-number, meaning that if mode-1 contains $n$ photons then mode-2 is in a state with exactly the same number of photons. The photon-number distiribution for the input state $P(n)= |\langle n | \psi \rangle|^2 $ can be calculated obtaining the well known sub-Poissonian quantum statistics of the from $P(n)=| \frac{\tanh{r}^{n}}{\cosh{r}}|^2$. Photon-number distributions for different values of the squeezing parameter $r=1, 0.5, 0.1$ are 
displayed in Fig. 1 (d), revealing the large vacuum contribution in 2-mode photon-number squeezed states. \\

\section{Quadrature Rotation}

The photon-number squeezed state depicted in Fig. 1 displays a standard Gaussian profile in quadrature space $(X,Y)$, with no topological charges or phase singularities. In order to imprint a vortex in quadrature space $(X,Y)$, 
we introduce a rotation $\hat{C}$ by an angle $\phi$, 
represented by a unitary operator of the form:

\begin{equation}
\hat{C}=e^{i2\phi [ \hat{a}^{\dagger}\hat{b} +\hat{b}^{\dagger}\hat{a}] },
\end{equation}

where $(\hat{a}^{\dagger},\hat{a})$ and $(\hat{b}^{\dagger},\hat{b})$ are creation and destruction operators for modes $(a,b)$, which satisfy the standard commutation rules $[\hat{a}^{\dagger},\hat{a}]=1$ and $[\hat{b}^{\dagger},\hat{b}]=1$. It can be readily seen that 
the operator $\hat{C}$ introduces a rotation around $z$-direction, since the OAM operator for photons in the Jordan-Schwinger representation can be written as $L_{z}= \hat{a}^{\dagger}\hat{b} +\hat{b}^{\dagger}\hat{a}$ \cite{Leuchs}, where $L_{z}$ is the generator of the rotations in $z$-direction.\\

The input state transformed under the unitary operator $\hat{C}$ becomes $|\psi' \rangle$ :

\begin{equation}
|\psi' \rangle = e^{i 2 \phi \hat{C}}| \psi \rangle, 
\end{equation}

which represents a rotation of the quadratures by an angle $\phi$. In the Heisenberg picture, considering standard commutation rules for creation and annhiliation operators, we obtain the following expression [see Appendix A]:

\begin{eqnarray}
|\psi' \rangle & = &\frac{D}{\cosh {r}} \sum_{j=0} ^{N/2} (\tanh{r})^{j} \times\\
& & (\frac{\hat{a}^{\dagger}}{\sqrt{2}}+ i \frac{\hat{b}^{\dagger}}{\sqrt{2}})^{j} (\frac{\hat{b}^{\dagger}}{\sqrt{2}}+ i \frac{\hat{a}^{\dagger}}{\sqrt{2}})^{j} |0,0\rangle. \notag
\end{eqnarray}

By a binomial expansion in Eq. (5) we obtain:

\begin{eqnarray}
|\psi' \rangle & = & \frac{D}{\cosh {r}} \sum_{j=0} ^{N/2} A^{r,N}_{j} \times \\
& & \sum_{k=0} ^{j} \sum_{l=0} ^{j} B^{\phi}_{k,l} C^{N j}_{k l} |j-(l-k), j+(l-k) \rangle. \notag
\end{eqnarray}

where $D$ is the normalization factor. The coefficients in the sums are of the form $A^{r,N}_{j}= (\tanh{r})^{j} \sqrt{\frac{j! (N-j)!}{2^{N}}} $, $B^{\phi}_{k,l}=( i 2 \phi)^{l+k}$, while $C^{N j}_{k,l}$ takes the form [see Appendix A]:

\begin{equation}
C^{N j}_{l k}= \frac{\sqrt{(N-j-l+k)!(j+l-k)!}}{k!(j-k)!l!(N-j-l)!}. 
\end{equation}

In order to observe the action of the rotation $\hat{C}$ in quadrature space we turn to the spatial representation of the transformed ket $\psi' (x,y)= \langle x,y | \psi' \rangle$.

\section{Laguerre-Gauss Mode Expansion} 

The spatial quadrature representation of the rotated state $\psi' (x,y)$ results in:

\begin{eqnarray}
\psi' (x,y) &= & \frac{D}{\cosh {r}} \sum_{j=0} ^{N/2} A^{r,N}_{j} \\
& & \sum_{k=0} ^{j} \sum_{l=0} ^{j} B^{\phi}_{k,l} C^{N j}_{l k} H_{j- (l-k)}(x) H_{j+(l-k)}(y)e^{-(x^2+y^2)/2}, \notag 
\end{eqnarray}

\begin{figure}[t!]
\includegraphics[width=0.4\textwidth]{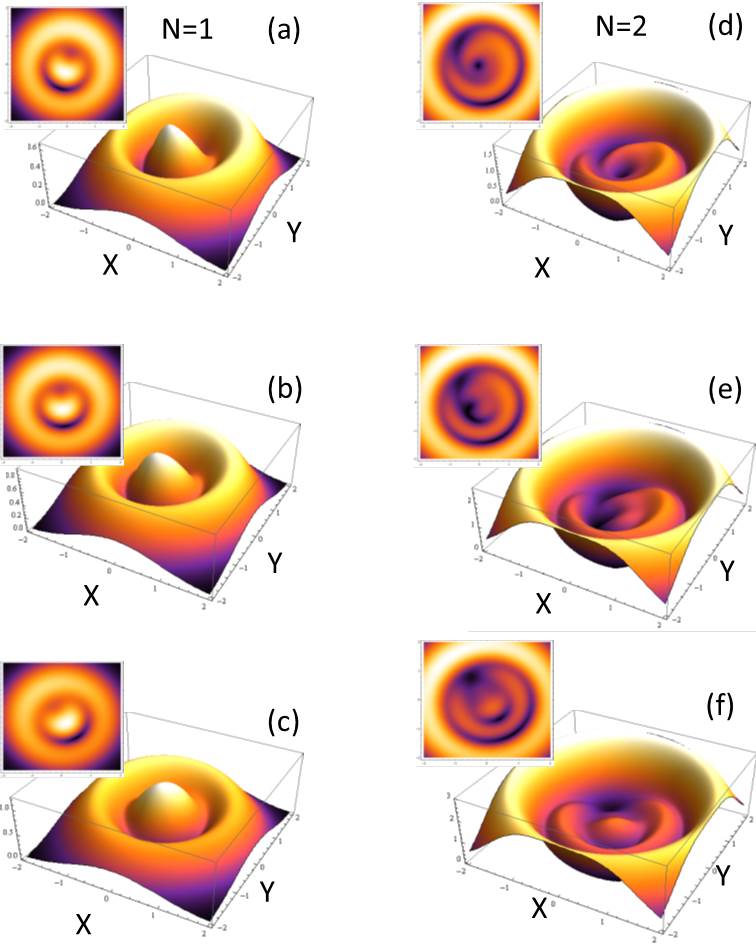}
\caption{3D plot of amplitude $|\psi_{LG}(x,y)|$ for resulting Laguerre-Gauss mode in quadrature space depicting the impact of the squeezing parameter $r$ in the formation of vortices in quadrature space, for different values of squeezing parameter $r$ and photon-number 
$N$. Insets correspond to contour plots of $|\psi_{LG}(x,y)|$. (a) $N=1$, $r=1$, (b) $N=1$, $r=0.5$, (c) $N=1$, $r=0.02$, (d) $N=2$, $r=1$, $N=2$, $r=0.5$, $N=2$, $r=0.02$. As the squeezing parameter decreases the formation of $N$ vortices becomes apparent (see text for details). }
\end{figure} 

\begin{figure*}[t!]
\centering
\includegraphics[width= 1\textwidth]{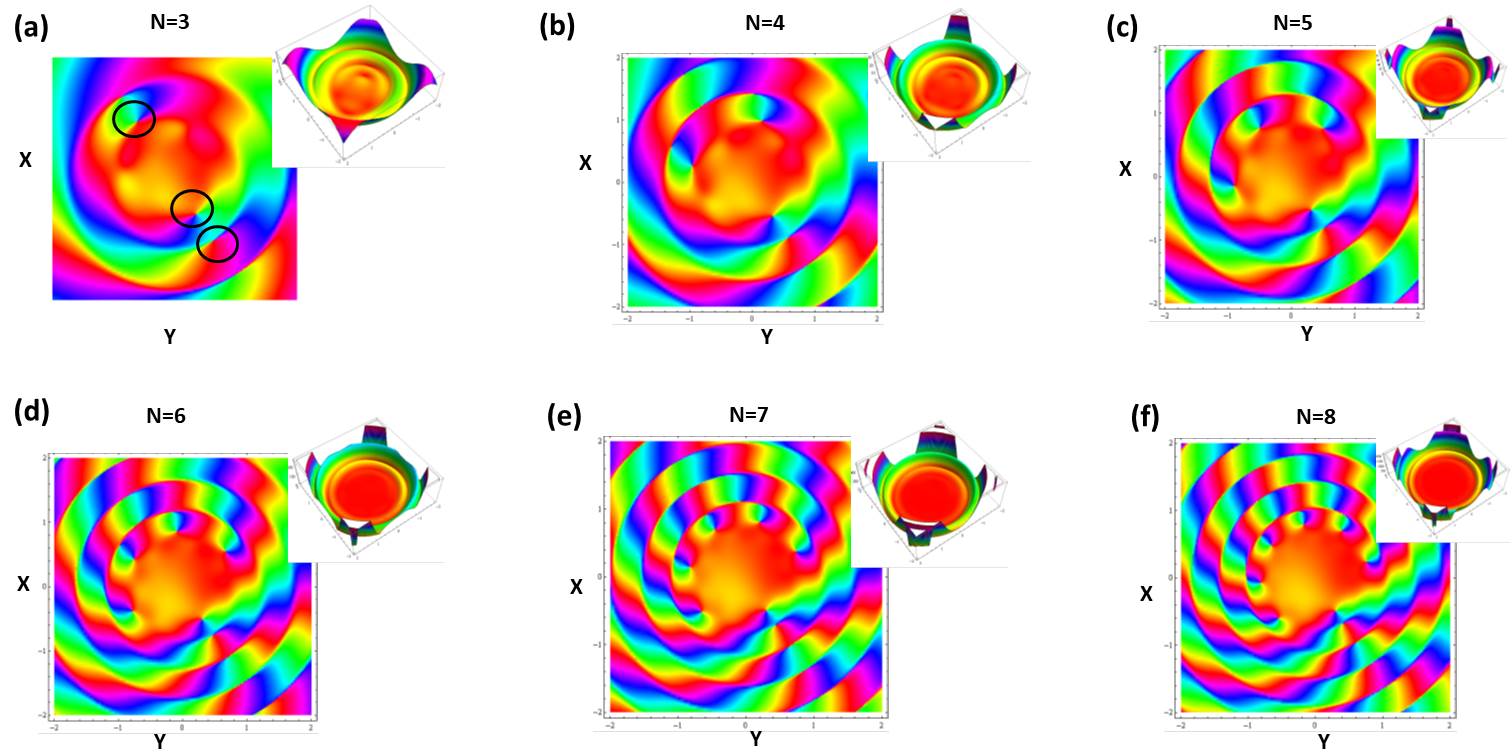}
\caption{Phase profile of resulting Laguerrre-Gauss mode $\psi_{LG}(x,y)$ for a squeezing parameter $r=0.02$, exploring the impact of the total photon-number $N$ in the formation of vortices in quadrature space. Insets correspond to amplitude plots $|\psi_{LG}(x,y)|$. The numerical results confirm creation of $N$ vortices for $N$ input photons. (a) $N=3$, (b) $N=4$, (c) $N=5$, (d) $N=6$, (e) $N=7$, (f) $N=8$ (see text for details).}
\end{figure*}

where $ H_{j- (l-k)}(x)=\langle x | j- l+k \rangle$ and $ H_{j +(l-k)}(y)=\langle y | j+l-k \rangle$ are Hermite-Gauss polynomials of order $(j-l+k)$ and $(j+l-k)$, respectively. \\

It is well known that Hermite-Gauss (HG) modes with spatial dependence $H_{p}(x)H_{q}(y)$ may become a single Laguerre-Gauss (LG) mode of order $L_{q}^{p-q}(x^2+y^2)$ provided a phase change of $\pi/2$ in the mode profile is achieved \cite{11}. 
Such Laguerre-Gauss mode is associated with a vortex number of $(p-q)$ [21-34].\\

By choosing the rotation parameter $\phi=\pi/4$, we may obtain the required phase change to convert the Hermite-Gauss modes into a single Laguerre-Gauss mode. By relabeling the indices $(l-k)=m$, with $m=0,...,N/2$, 
we note the spatial quadrature profile can be written as a sum of products of
HG modes of the form $H_{j-m}(x)H_{j+m}(y)$. Selecting $\phi=\pi/4$, the spatial quadrature profile can be written in terms of LG modes of the form $L_{j-m}^{2m}(r^2)$, thus resulting in a superposition of LG modes of order $2m$. 

\section{Numerical Results}

To explore the resulting mode-profile in quadrature space $(X,Y)$, we performed numerical simulations for a superposition of LG modes of the form:
\begin{equation}
\psi_{LG}(x,y)= \frac{1}{\cosh{r}}\sum_{j=0}^{N/2}\sum_{m=0}^{j}A^{r,N}_{j} L_{j-m}^{2m}(x^2+y^2)e^{-(x^2+y^2)/2},
\end{equation}

where $r$ is the squeezing parameter and the coefficients take the form $A^{r,N}_{j}= (\tanh{r})^{j} \sqrt{\frac{j! (N-j)!}{2^{N}}} $.\\

We performed numerical simulations in quadrature space for different values of squeezing parameter $r$, and different values of photon-number $N$. The results are depicted in Fig. 2 and Fig. 3. 
The main result we observe is that, for a sufficiently small squeezing parameter $r$, the resulting spatial profile exhibits $N$ vortices for an input state with $N$ photons, as expected for a Laguerre-Gauss mode of order $2m$ with $m=N/2$. In this way, we have 
mapped the reduced uncertainty in photon-number in Fock space, to a reduced uncertainty in vortex-number in quadrature space. 

\subsection{Dependence on squeezing parameter $r$}

In order to better understand the impact of the squeezing parameter $r$ in the formation of vortices in quadrature space, we performed numerical simulations for different squeezing parameters, and for different number of photons. This is displayed in Fig. 2 (a)-(f). Fig. 2 left column corresponds to $N=1$ photon and right column corresponds to $N=2$ photons. Different rows in 
in decreasing order correspond to squeezing parameters $r=1, 0.5, 0.02$. Numerical simulations clearly reveal that vortices are formed as $r$ decreases, thus as the uncertainty in photon-number decreases as expected. Thus confirming that the reduced uncertainty in Fock space is mapped to reduced uncertainty in vortex number, in quadrature space.

\subsection{Dependence on photon-number $N$}

To confirm the viability of generation of high-order vortex states in quadrature space we performed numerical simulations for larger total number of photons ($N>2$). This is depicted in Figure 3, for a squeezing parameter $r=0.02$. Fig. 3 (a)-(f) display plots of phase profile associated with $\psi_{LG}(x,y)$, calculated via $\phi=\tan^{-1}[\frac{\Im[\psi_{LG}(x,y)]}{\Re[\psi_{LG}(x,y)]}]$, for $N=3,4,5,6,7,8$ photons in the input 2-mode state, further confirming the azymuthal charge and vorticity in quadrture space increases with the number of photons $N$. Insets display 3D plots of mode amplitude $|\psi_{LG}(x,y)|$. 
As predicted, in all cases the number of vortices in quadrature space is equal to the total number of photons $N$ in the initial 2-mode photon-number squeezed state, thus confirming the mapping of photon-number in Fock space to vortex-number in quadrature space.

\section{Photon-number distribution of vortex-squeezed states}

The generation of vortices in quadrature space can be considered an interference effect arising from photon-number fluctuations, therefore it is expected that the photon-number distribution should be modified for vortex states. 
To further confirm that photon-number fluctuations are mapped into interference effects in quadrature space, resulting in the emergence of $N$ vortices, for an $N$ truncated photon-number squeezed input state, 
we calculated the photon-number distribution for the vortex states $P(n_{1},n_{2})=|\langle n_{1},n_{2}|\psi' \rangle|^2$. Using orthogonality of Fock states, the only terms that survive are $j=\frac{n_1+n_2}{2}$ and $l=k +\frac{n_1-n_2}{2}$. 
The sums in Eq. (6) collapse into a single sum of the form:

\begin{equation}
P(n_{1},n_{2})=|\frac{D}{\cosh{r}}\sum_{k=0}^{\frac{n_1+n_2}{2}}A_{n_{1},n_{2}}^{N,r} B_{n_{1},n_{2}}^{\phi,k}C_{n_{1},n_{2}}^{N,k}|^2,
\end{equation}

where $A_{n_{1},n_{2}}^{N,r}=\tanh(r)^{\frac{n_1+n_2}{2}} \sqrt{\frac{(n_1+n_2)!(N-n_1+n_2)!}{2^{N}}}$, $B_{n_{1},n_{2}}^{\phi,k}=(i 2 \phi)^{2k + \frac{n_1-n_2}{2}}$, and $C_{n_{1},n_{2}}^{N,k}$ results in:

\begin{equation}
C_{n_{1},n_{2}}^{N,k}=\frac{\sqrt{(N-n2)!(n_2!)}}{\sqrt{k! (\frac{n_1+n_2}{2}-k)!(N-n_2-k)!}}.
\end{equation}

Eq. (11) reveals the photon-number fluctuations which give rise to the emergence of vortices in quadrature space. Numerical results for the photon-number distributions of vortex states are presented in Fig. 4, confirming the predicted photon-number fluctuations and super-Poissonian statistics.

\begin{figure}[h]
\hspace{-0.5cm}
\includegraphics[width=0.5\textwidth]{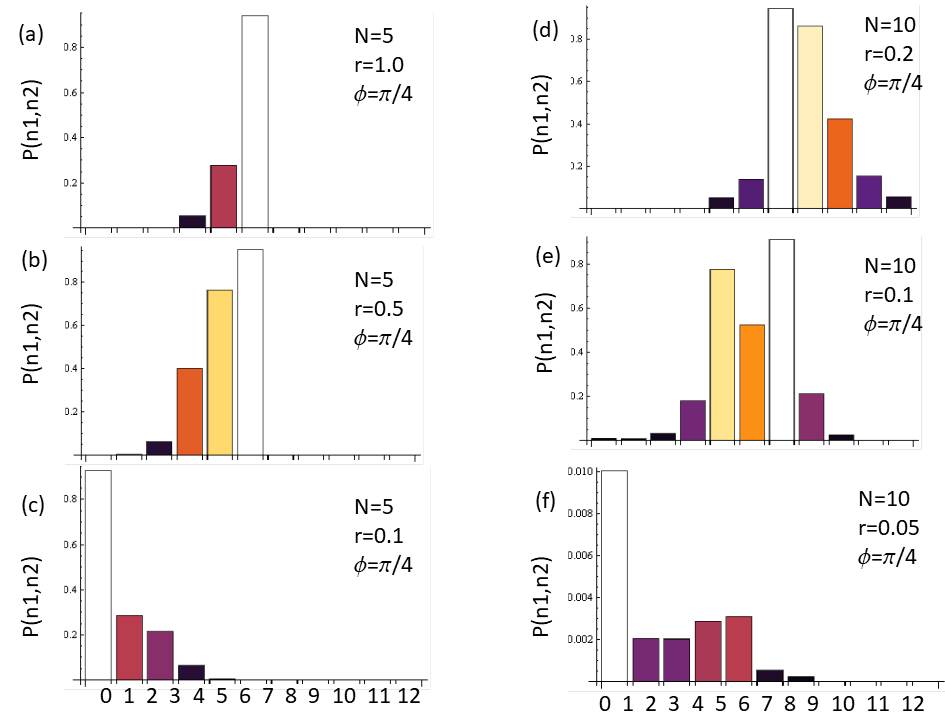}
\caption{Photon-number statistics $P(n_1,n_2)$ for vortex states considering $n_1=n_2$. Left column $N=5$ and $\phi=\pi/4$, right column $N=10$ and $\phi=\pi/4$. Different rows correspond to squeezing parameters (a) $ r=1.0$, (b) $ r=0.5$, (c)$ r=0.1$, (d) $r= 0.2$, (e)$ r=0.1$, (f) $r= 0.05$. The photon-number fluctuations due to vortex formation is revealed (see text for details). }
\end{figure}

\section{Discussion}

We presented a scheme for generation of high-order vortex states starting from a two-mode photon-number squeezed state generated via the non-linear process of Spontaneous Parametric Down Conversion (SPDC). 
By applying a parametric rotation in quadrature space $(X,Y)$ using a $\phi$ converter, the quadrature representation of the photon-number squeezed input state is transformed into a high-order vortex state, with $N$ vortices,
for a input state containing $N$ photons, thus mapping the fluctuations in photon-number to interference effects in quadrature space as depicted by optical singularities with zero-point intensity. Furthermore, we obtained analytical and numerical expressions for 
the super-Poissonian photon-number statistics and fluctuations, giving rise to vortex formation in quadrature space. Our scheme has the potential of exploiting the advantages of optical vortices, such as ease of creation and detection, high dimensionality and topological properties, for quantum applications requiring squeezed uncertainty beyond the SQL limit ($\sqrt{N}$), such as quantum cryptography, quantum metrology and quantum sensing [35-40].

\bigskip

\section{Acknowledgements}
The authors acknowledge A. Aiello and A. Lvovsky for helpful discussions. GP acknowledges financial support via grants PICT Startup 2015 0710, and UBACyT PDE 2017.

\section{Appendix A}

The starting point of the derivation is Eq. 5, which defines a $\pi/4$ mode converter: 

\begin{equation}
\label{Eq1:Mode_Converter}
\hat{C} = \frac{1}{2}\left(a^{\dagger}b + ab^{\dagger}\right)
\end{equation}

\noindent where $a \left(b\right)$ are the bosonic mode operators acting on orthogonal modes and follow regular bosonic commutation relations. Also, let us consider the initial state of the two mode system to be the following

\begin{equation}
\label{Eq2:Initial_State}
\vert\psi\rangle = \sum_j A_j \vert N-j\rangle_a \vert j \rangle_b
\end{equation}

\noindent The above describes a general two-mode state in the Fock basis with total number of particles $N$ distributed between the two modes. Now Eq. \ref{Eq2:Initial_State} can be written in terms of the mode operators as follows

\begin{equation}
\label{Eq3:Intial_State_Mode_Operators}
\vert\psi\rangle = \sum_j A_j \left(a^{\dagger}\right)^{N-j}\left(b^{\dagger}\right)^{j} \vert 0 \rangle_a \vert 0 \rangle_b
\end{equation}

\noindent where it is understood that the operator $a \left(b\right)$ acts on mode $\vert 0 \rangle_a \left(\vert 0 \rangle_b \right)$. We want to find how the state $\vert \psi \rangle$ transforms under the action of $\hat{C}$. Moving to the Heisenberg picture, the mode operators $a \left(b\right)$ evolve under $\hat{C}$ as

\begin{equation}
\label{Eq4:Operator_Evolution}
a \rightarrow \exp\left(i2\phi\hat{C}\right) a \exp\left(-i2\phi\hat{C}^{\dagger}\right)
\end{equation}

Using the Baker-Hausdorff lemma, we can write Eq. \ref{Eq4:Operator_Evolution} as follows

\begin{eqnarray}
\label{Eq5:BCH_Expansion}
\exp\left(i2\phi\hat{C}\right) a \exp\left(-i2\phi\hat{C}^{\dagger}\right) & = & a + i2\phi\left[\hat{C},a\right] + \\
& & \frac{\left(i2\phi\right)^2}{2!}\left[\hat{C},\left[\hat{C},a\right]\right] + ... \notag \\
\end{eqnarray}

Solving for the commutators, we see that $\left[\hat{C},a\right] = -b/2$ and $\left[\hat{C},b\right] = -a/2$. Plugging these values back into Eq. \ref{Eq5:BCH_Expansion}, we see that we can group the terms as

\begin{eqnarray}
\label{Eq6:Grouping_together}
a\left(1 - \frac{\left(\phi\right)^2}{2!} +...\right)-ib\left(\phi - \frac{\phi^3}{3!} +...\right)\nonumber \\
= a\cos{\phi} - ib\sin{\phi}
\end{eqnarray}

Now for a $\pi/4$ mode converter, we put $\phi = \pi/4$ resulting in the transformation

\begin{equation}
\label{Eq7:Final_transformation}
a \rightarrow \frac{1}{\sqrt{2}}\left(a - ib\right)
\end{equation}

and similarly for $b$. Therefore, Eq. \ref{Eq3:Intial_State_Mode_Operators} is transformed to 

\begin{equation}
\label{Eq8:Final_state}
\vert \psi \rangle_v = \sum_j A_j \left(a^{\dagger}+ib^{\dagger}\right)^{N-j}\left(b^{\dagger}+ia^{\dagger}\right)^{j} \vert 0 \rangle_a \vert 0 \rangle_b \notag
\end{equation}

\noindent under the effect of the $\pi/4$ mode converter. It is understood in Eq. \ref{Eq8:Final_state} that factors $1/\sqrt{2}$ have been absorbed into $A_j$. Eq. 6 follows from here by a binomial expansion of the terms.

\end{document}